\documentclass[prb,twocolumn,showpacs,superscriptaddress,preprintnumbers]{revtex4}

\usepackage{graphicx}
\usepackage{dcolumn}
\usepackage{amsmath}
\usepackage{color}

\begin{document}

\title{Magnetic Collapse and the Behavior of Transition Metal Oxides at High Pressure} 

\author{I. Leonov}
\affiliation{Theoretical Physics III, Center for Electronic Correlations and
Magnetism, Institute of Physics, University of Augsburg, 86135 Augsburg, Germany}

\affiliation{Materials Modeling and Development Laboratory, National University of Science and Technology 'MISIS', 119049 Moscow, Russia}

\author{L. Pourovskii}
\affiliation{Centre de Physique Th\'eorique, \'Ecole Polytechnique, CNRS, Universit\'e Paris-Saclay, 91128 Palaiseau, France}

\affiliation{Coll\`ege de France, 11 place Marcelin Berthelot, 75005 Paris, France}

\affiliation{Materials Modeling and Development Laboratory, National University of Science and Technology 'MISIS', 119049 Moscow, Russia}

\author{A. Georges}
\affiliation{Centre de Physique Th\'eorique, \'Ecole Polytechnique, CNRS, Universit\'e Paris-Saclay, 91128 Palaiseau, France}

\affiliation{Coll\`ege de France, 11 place Marcelin Berthelot, 75005 Paris, France}

\affiliation{DQMP, Universit\'e de Gen\`eve, 24 quai Ernest Ansermet, CH-1211 Gen\`eve, Suisse}

\author{I. A. Abrikosov}
\affiliation{Department of Physics, Chemistry and Biology (IFM), Link\"oping University, SE-58183 Link\"oping, Sweden}
\affiliation{Materials Modeling and Development Laboratory, National University of Science and Technology 'MISIS', 119049 Moscow, Russia}

\date{\today}

\begin{abstract}
We report a detail theoretical study of the electronic structure and phase stability of
transition metal oxides MnO, FeO, CoO, and NiO in their paramagnetic cubic B1 structure 
by employing dynamical mean-field theory of correlated electrons combined with \emph{ab initio} band structure methods (DFT+DMFT). Our calculations reveal that under pressure these materials exhibit a Mott insulator-metal transition (IMT) which is accompanied by a simultaneous collapse of local magnetic moments and lattice volume, implying a complex interplay between chemical bonding and electronic correlations. Moreover, our results for the transition pressure show a monotonous decrease from $\sim 145$ GPa to 40 GPa, upon moving from MnO to CoO. In contrast to that, in NiO, magnetic collapse is found to occur at remarkably higher pressure of $\sim 429$ GPa. 
We provide a unified picture of such a behavior and suggest that it is primary a localized to itinerant moment behavior transition at the IMT that gives rise to magnetic collapse in transition metal oxides.

\end{abstract}

\pacs{71.10.-w, 71.27.+a, 71.30.+h} \maketitle

%%%%%%%%%%%%%%%
% Introduction
%%%%%%%%%%%%%%%

The effect of strong electron correlations plays a crucial role in the formation of a variety of electronic and magnetic phenomena experimentally observed in transition metal oxides \cite{Review}. The pressure-induced Mott insulator-metal transition (IMT) \cite{RXP14} is of particular importance for both fundamental science and modern technology; for example, due to its potential application in micro- and opto- electronics. Moreover transition metals Mn, Co, and Ni are known to be siderophile elements, i.e., together with iron are extensively abundant in the Earth's interior \citep{Siderphile}. Because of that, the electronic structure and phase stability of these basic oxide compounds, especially, in the high-pressure and temperature range, is of fundamental importance for understanding the properties and evolution of the Earth's lower mantle \cite{Lower_mantle}.

It has been shown that the pressure-induced Mott insulator-metal transition in these compounds is followed by a magnetic collapse, a remarkable reduction of magnetic moments of transition metal ions \cite{IC93,HS-LS+MIT}, which strongly influences their electronic and structural properties.  
Since our present knowledge of the Earth's interior structure is mainly based on average geophysical observations, e.g., from seismology, quantifying the effects of magnetic collapse on the electronic and structural properties of these minerals is of particular interest for geophysics. 
%
%In fact, variations of density and elastic moduli control the speed of seismic waves, i.e., play a crucial role in the interpretation of seismological data \cite{Lower_mantle}.
%
While the properties can be calculated employing, e.g., standard band-structure methods, these techniques cannot capture all generic aspects of a Mott IMT, such as a formation of the lower and upper-Hubbard bands, coherent quasiparticle behavior, strong renormalization of the electron mass, etc., because of neglecting the effect of strong correlations of localized $3d$ electrons. Therefore these methods have only limited applicability and cannot provide a quantitative description of the electronic and lattice properties. In particular, at low pressures they predict metallic behavior for FeO and CoO, which are, in fact, insulators \cite{IC93,TOW84,AZA91}.
This obstacle can be overcome by employing, e.g., a state-of-the-art method for calculating the electronic structure of strongly correlated materials (DFT+DMFT) \cite{DMFT,LDA+DMFT}.
It merges \emph{ab initio} band-structure techniques, such as the local density approximation (LDA) or the generalized gradient approximation (GGA), with dynamical mean-field theory (DMFT) of correlated electrons \cite{DMFT}, providing a good quantitative description of the electronic and lattice properties \cite{SK01,MH03,KH04,KK09,LB08,LP11,GP13,SH06,AP11,PM14,LA14,KL08,SP10,OC12,IL15,HW12,RL06,BK12}.
In particular, this advance theory makes it possible to determine the electronic structure and phase stability of paramagnetic correlated materials at finite temperatures, e.g., near a Mott insulator-metal transition \cite{KH04,KK09,LB08,PM14,LA14,KL08,SP10,OC12,IL15,HW12,RL06,BK12}.
The DFT+DMFT approach has been used to study the electronic and structural properties of correlated electron materials \cite{SK01,MH03,KH04,KK09,LB08,LP11,GP13,SH06,AP11,PM14}, including transition metal monoxides \cite{KL08,SP10,OC12,IL15,HW12,RL06,BK12,NP12}. 
% 
%Nevertheless, in spite of long-term intensive research, the high-pressure behavior of transition metal oxides, as well as several key properties, such as electronic structure, magnetic state and phase stability, are still poorly understood.
%
In practice however these calculations employed different approximations, resulting in various scenarios for the IMT. For example, a local moment collapse at the IMT was found to occur in MnO \cite{KL08}; for FeO different calculations predicted either a Mott transition in the high-spin state followed by a slow crossover into the low-spin one \cite{OC12} or the absence of magnetic collapse at all \cite{SP10}.
In addition, most of these theoretical calculations used experimental equations of state, neglecting the effects of coupling between the electrons and lattice at the IMT \cite{SP10,OC12,HW12}. %The self-consistency in the charge density was not taken into account, apart from the work of Ohta {\it et al.}.~\cite{SP10} and Leonov. 

In this work, we employ a state-of-the-art fully charge self-consistent DFT+DMFT approach \cite{charge-sc-LDA+DMFT} to explore the electronic structure, magnetic state, and phase stability of all late transition metal monoxides, from MnO to NiO.  We use this advanced theory to systematically study these oxides at high pressure and temperature in their paramagnetic cubic B1 lattice structure.
We find that in all four compounds the Mott insulator-metal transition is accompanied by a simultaneous collapse of local moments. Our results suggest that the magnetic collapse is driven by a transition from the localized to itinerant-moment behavior at the IMT. The phase transition is of the first order with a significant fractional volume collapse $\Delta V/V$ ranging from 8.6 \% to 13.6 \% for MnO, FeO and CoO, and it is only about 1.4 \% for NiO. 
We predict a monotonous decrease of the transition pressure from MnO to CoO, namely, from $\sim 145$ GPa to 40 GPa, implying a complex interplay between chemical bonding and electronic correlations in these materials. In contrast to that, the Mott transition and magnetic collapse is predicted to occur at $\sim 429$ GPa in NiO, i.e., has a substantially higher value than that in the other monoxides, which can be understood as a crossover in the effective degeneracy of low-energy excitations from five-orbital (as in MnO, FeO, and CoO) to two-orbital behavior (as in NiO). 
Overall, our calculations for the first time provide a unified picture for the pressure evolution of all late transition metal monoxides, that of a first-order Mott transition accompanied by a simultaneous collapse of local magnetic moments.
%
%Our results clearly indicate the crucial importance of electronic correlations to explain the electronic structure and lattice properties of transition metal oxides.

%%%%%%%%%%%%%%%%%%%%%%
% Equation of states
%%%%%%%%%%%%%%%%%%%%%%

We here investigate the electronic and structural properties of correlated transition metal oxides MnO, FeO, CoO, and NiO at high pressure and temperature using the GGA+DMFT computational approach (GGA: generalized gradient approximation). 
To this end, we employ a fully charge self-consistent DFT+DMFT scheme \cite{charge-sc-LDA+DMFT} implemented with plane-wave pseudopotentials \cite{Pseudo} to calculate the electronic structure, magnetic state, and phase stability of these compounds. 
For the partially filled Mn, Fe, Co, and Ni $3d$ and O $2p$ orbitals we construct a basis set of atomic-centered symmetry-constrained Wannier functions \cite{MV97,Wannier-functions}. To solve the realistic many-body problem, we employ the continuous-time hybridization-expansion quantum Monte-Carlo algorithm \cite{ctqmc}. The calculations are performed for a cubic rocksalt (B1) crystal structure in the paramagnetic state at temperature $T = 1160$ K. 
We use the following values of the average Hubbard $U$ and Hund's exchange $J$ as estimated \cite{AZA91,KL08,SP10,OC12,IL15,HW12,RL06,KA10} previously: $U=8.0$ eV and $J=0.86$ eV for the Mn $3d$ orbitals, 7.0 eV and 0.89 eV for Fe, 8.0 eV and 0.9 eV for Co, and 10.0 eV and 1.0 eV for Ni, respectively. The Coulomb interaction has been treated in the density-density approximation. The spin-orbit coupling was neglected in these calculations.
Moreover, the $U$ and $J$ values are assumed to remain constant upon variation of the lattice volume. We employ the fully-localized double counting correction, evaluated from the self-consistently determined local occupations, to account for the electronic interactions already described by GGA. The spectral functions were computed using the maximum entropy method.%

\begin{figure}[tbp!]
\centerline{\includegraphics[width=0.5\textwidth,clip=true]{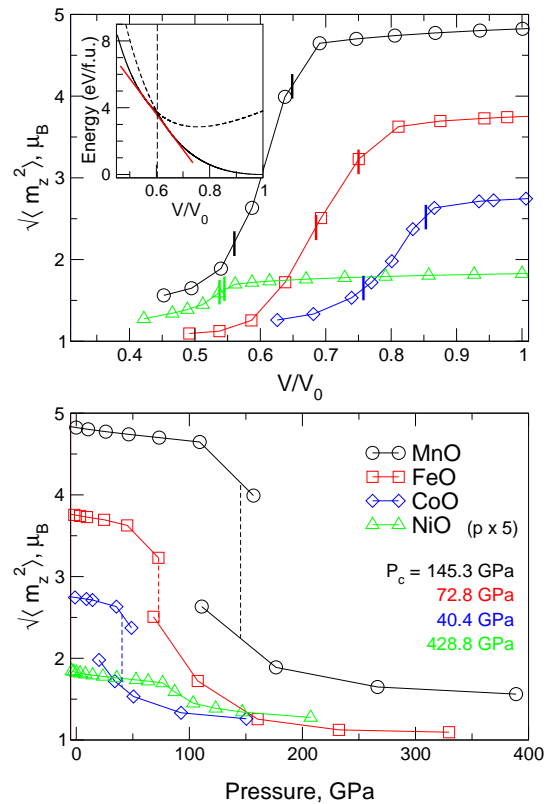}}
\caption{Local moment $\sqrt{\langle m_z^2 \rangle}$ calculated by DFT+DMFT as a function of lattice volume (top) and pressure (bottom). The IMT associated with the lattice volume collapse is shown by vertical solid (top) and dashed lines (bottom). The inset shows the total energy calculated by DFT+DMFT and the Maxwell construction for paramagnetic MnO. Note that for NiO the pressure scale is shrunk by factor 5.
}
\label{Fig_1}
\end{figure}

\begin{figure}[tbp!]
\centerline{\includegraphics[width=0.5\textwidth,clip=true]{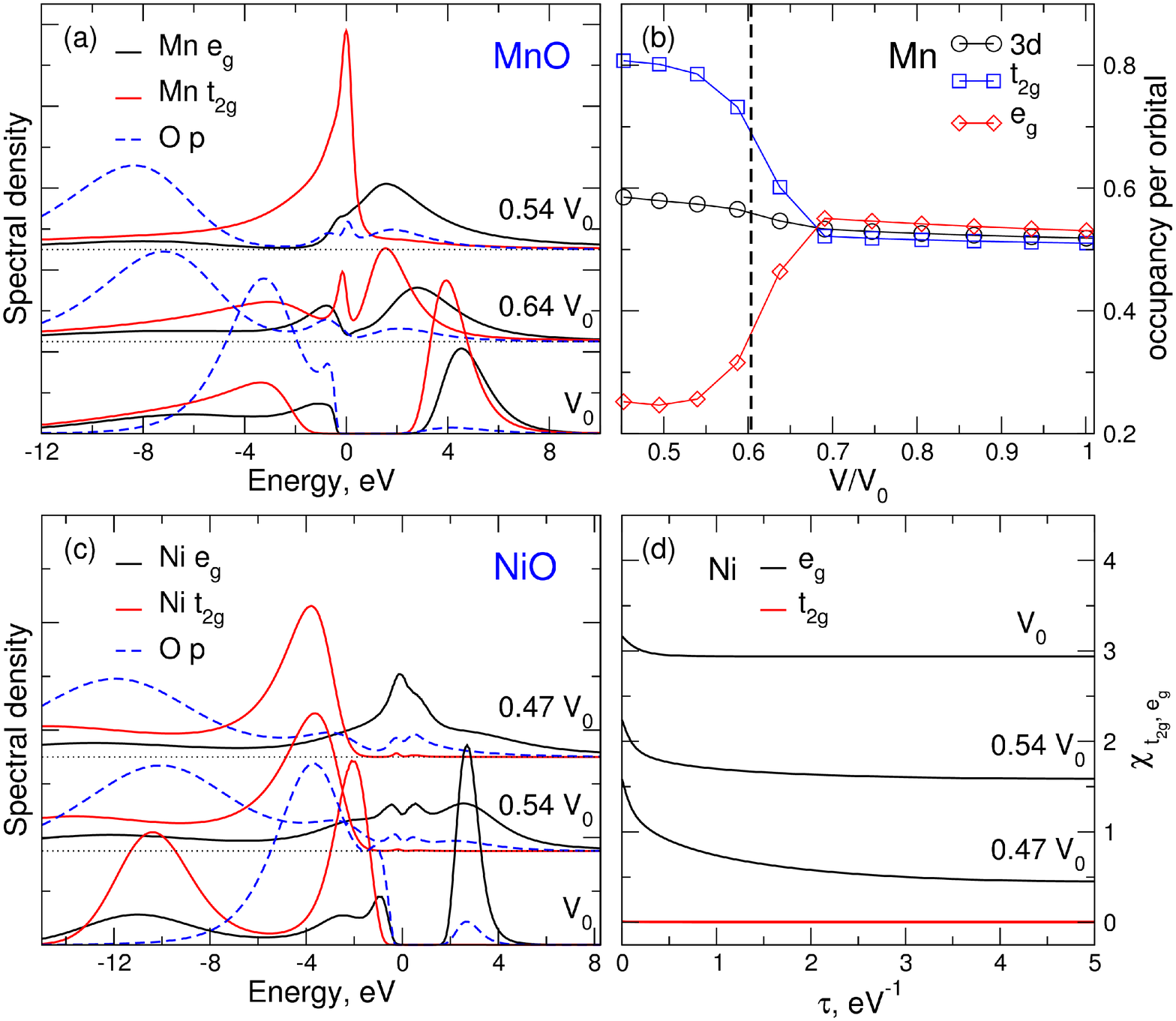}}

\caption{
Evolution of the spectral function (a) and orbital occupations (b) calculated by DFT+DMFT for paramagnetic MnO as a function of volume. Spectral function (c) and local spin-spin correlation function $\chi(\tau)=\langle \hat m_z(\tau)\hat m_z(0) \rangle$ (d) of paramagnetic NiO. $\tau$ is the imaginary time. 
At equilibrium volume $V_0$, the $3d$ electrons are localized to form fluctuating moments [$\chi(\tau)$ is seen to be almost constant and close to its maximal value $S=1$ for the Ni $e_{g}$ states]. At high compression, the $3d$ electrons show an itinerant magnetic behavior, implying a localized to itinerant moment crossover under pressure.
}
\label{Fig_2}
\end{figure}

Our results for the local-moment vs. volume and pressure equations of state are summarized in Fig. \ref{Fig_1}.
%
%%%%%%%%%%%%%%%%%%%%%%%%%%%%%%%%%%%%%%%%%%%%%%%%%%%%%%%%%%%%%%%%%%%%%%%%%%%%%%%%%
%%%%%%%%%%%%%%%%%%%%%%%%%%%%%%%%%%%%%%%%%%%%%%%%%%%%%%%%%%%%%%%%%%%%%%%%%%%%%%%%%
%
To this end, we calculate the total energy and fluctuating (instantaneous) local moment $\sqrt{\langle m_z^2 \rangle}$ for these compounds as a function of volume.
We evaluate pressure by fitting our results for the total energy to the third-order Birch-Murnaghan equation of states separately for the low- and high-volume regions.
We obtain that all four compounds exhibit magnetic collapse -- a remarkable reduction of the local magnetic moments upon compression of the lattice.
Indeed, the local moments are seen to retain their high-spin values, e.g., for NiO and MnO of $\sim 1.8$ to 4.8 $\mu_B$, respectively, upon compression down to about $0.6-0.7$ $V/V_0$.
In fact, these values correspond to the high-spin magnetic state of the Mn$^{2+}$ and Ni$^{2+}$ ions. Thus, in a cubic crystal field, the Mn$^{2+}$ ($3d^5$ electronic configuration) and Ni$^{2+}$ ($3d^8$) ions have a local moment of 5 $\mu_B$ and 2 $\mu_B$, respectively.
Upon further compression, a high-spin (HS) to low-spin (LS) crossover takes place in all four materials, with a collapse of the local moments to a LS state. The resulting low-spin local moments are about 1.6, 1.1, 1.26, and 1.28 $\mu_B$ for MnO, FeO, CoO, and NiO, respectively.
In addition, upon compression, a substantial redistribution of electrons between the $t_{2g}$ and $e_g$ orbitals is found within the transition metal $3d$ bands of MnO, FeO, and CoO (shown for MnO in Fig. \ref{Fig_2}(b)). Thus, the $t_{2g}$ orbital occupations are found to gradually increase with pressure, whereas the $e_g$ orbitals are strongly depopulated (below 0.27 for MnO and FeO; and 0.44 for CoO). The $3d$ total occupancy remains essentially unchanged with pressure. We therefore interpret this spin crossover as a HS-LS transition.

%%%%%%%%%%%%%%%%%%%%%%%%%%%%%%%%
% Magnetic transition pressures
%%%%%%%%%%%%%%%%%%%%%%%%%%%%%%%%

\begin{table}
\centering
\begin{tabular}{lccccc} 
\hline
Oxide & $P_{\rm c}$ (GPa) & $V_0$ (a.u.$^3$) & $V_{\rm tr}$ (a.u.$^3$) & $\Delta V/V$ (\%) & $K$ (GPa) \\
\hline
MnO & 145.3 & 158.9 & 103.1 /  89.0 & 13.6 & 137 / 263  \\
FeO & 72.8  & 146.3 & 109.7 / 100.2 & 8.6  & 140 / 162  \\
CoO & 40.4  & 137   & 116.9 / 103.8 & 11.2 & 184 / 246  \\
NiO & 428.8 & 128   &  69.8 /  68.8 & 1.4  & 187 / 188  \\
\hline
\end{tabular}

\caption{Calculated structural parameters for the paramagnetic B1 phase of transition metal oxides. V$_0$ is ambient pressure volume; $V_{\rm tr}$ are the lattice volume collapse values; $K$, bulk modulus for the low/high pressure phase; $K' \equiv dK/dP$ is 4.1 for MnO, FeO, and CoO; $K' = 4.3$ for NiO.}

\label{tab:Table1}
\end{table}

Our results reveal that the magnetic transition pressures vary substantially among all these compounds. In particular, for MnO, magnetic collapse is found to take place at relatively high pressure of about 145 GPa (see Table \ref{tab:Table1}) \cite{different_U_J}.
Moreover, for FeO and CoO, it occurs at remarkably smaller values of about 73 GPa and 40 GPa, respectively.
That is, the transition pressure appears to monotonously decrease from MnO to CoO, which implies a complex interplay between chemical bonding and electronic correlations in these materials.
However, in contrast to that, NiO has a high transition pressure $\sim 429$ GPa, which is considerably larger than that of MnO, FeO, and CoO.
This remarkable behavior can be understood as a continuous reduction of the strength of electronic correlations upon changing of the electron configuration from $3d^5$ in Mn$^{2+}$ ions to $3d^7$ in Co$^{2+}$.
In fact, the effective interaction strength $U_{\rm eff}=d^{n+1}+d^{n-1}-2d^n$ and, hence, the tendency towards localization of the $3d$ electrons, changes from $U+4J$ for MnO to $U-3J$ for CoO. On the other hand, it increases substantially in NiO due a crossover in the effective degeneracy of low-energy excitations from five-orbital (as in MnO, FeO, and CoO) to two-orbital behavior (as in NiO).

%%%%%%%%%%%%%%%%%%%%%
% Spectral properties
%%%%%%%%%%%%%%%%%%%%%

\begin{figure}[tbp!]
\centerline{\includegraphics[width=0.5\textwidth,clip=true]{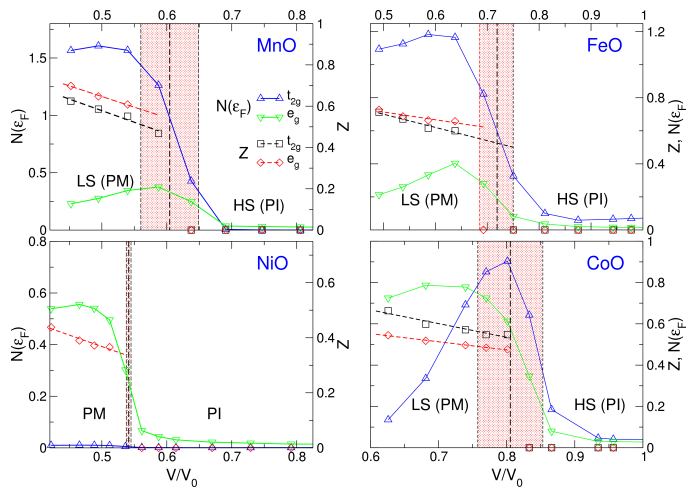}}
\caption{
Quasiparticle weight and spectral weight at the Fermi level calculated for the $t_{2g}$ and $e_g$ orbitals as a function of lattice volume. Our result for the lattice volume collapse calculated by DFT+DMFT is marked by a red shaded rectangle. The quasiparticle weight $Z=[1 - \partial Im \Sigma(i\omega)/\partial i\omega]^{-1}$ is evaluated from the slope of the polynomial fit of the imaginary part of the self-energy $\Sigma(i\omega_n)$ at $\omega=0$. 
}
\label{Fig_3}
\end{figure}

Next we address the spectral properties of paramagnetic transition metal oxides. In Fig. \ref{Fig_2}(a, c) we present our results for the evolution of the spectral function of MnO and NiO as a function of lattice volume. 
We note that at ambient pressure all these compounds (from MnO to NiO) are Mott insulators with large $d$-$d$ energy gaps, in agreement with experiment.
This result is in remarkable contrast to that previously obtained by standard band-structure methods, which predict metallic behavior, e.g., for FeO and CoO. 
This implies the crucial importance of strong correlations of localized $3d$ electrons, to determine the electronic and magnetic properties of transition metal oxides.
Indeed, at ambient pressure, the $3d$ electrons are strongly localized as it is seen from the our result for the local spin susceptibility $\chi(\tau)=\langle \hat m_z(\tau)\hat m_z(0) \rangle$ (where $\tau$ is the imaginary time; see Fig. \ref{Fig_2}(d)). Upon further compression, the $3d$ electrons exhibit a crossover from localized to itinerant moment behavior which is associated with a Mott transition, as it is clearly seen in paramagnetic NiO.
Upon the Mott transition, all these materials exhibit a strongly correlated metallic behavior with formation of the lower and upper-Hubbard bands, and the quasiparticle peak at the Fermi level, associated with a substantial renormalization of the electron mass (see Fig. \ref{Fig_3}).
%

%%%%%%%%%%%%%%%%%%%%%%%%%%%%%%
%  and structural collapse
%%%%%%%%%%%%%%%%%%%%%%%%%%%%%%

We find that in all four compounds magnetic collapse is accompanied by a lattice collapse, which implies a complex interplay between electronic and lattice degrees of freedom at the transition. The phase transition is of first-order with significant from 8.6 to 13.6 \% fractional volume collapse, $\Delta V/V$, except for NiO, where a resulting change of the lattice volume is about $1.4$ \% (see Table \ref{tab:Table1}).
We note however that these values should be considered as an upper-bound estimate because of neglecting by multiple intermediate-phase transitions upon fitting the total-energy result to the third-order Birch-Murnaghan equation of states.
These findings may have important implications for geophysics. In particular, magnetic collapse should result in an anomalous behavior, e.g., of density and elasticity. On the other hand, variations of density and elastic moduli control the speed of seismic waves, playing a crucial role in the interpretation of seismological data \cite{Lower_mantle}.
Our results for the bulk modulus (see Table \ref{tab:Table1}) in the HS phase are remarkably smaller than that in the LS phase in all these compounds, implying an enhancement of the compressibility at the phase transition. Thus, we obtain that the high-pressure phase is denser and less compressible and thereby has a higher bulk sound velocity $\nu = (K/ \rho)^{1/2}$. %For example, for FeO, the calculated change of the bulk velocity at the magnetic collapse transition is about 2.8 \%.

In addition, we evaluated the quasiparticle weights by using a polynomial fit of the imaginary part of the self-energy $\Sigma(i\omega_n)$ at the lowest Matsubara frequencies $\omega_n$ (shown in Fig. \ref{Fig_3}).
%
%It is obtained as $Z = [1 - \d Im \Sigma(i\omega) / i\omega ]^{-1}$ %from the slope of the polynomial fit at $\omega = 0$.
%
These results suggest that magnetic collapse is accompanied by a simultaneous Mott metal-insulator transition and collapse of the lattice volume. The electronic effective mass diverges at the IMT, in accordance with a Brinkman-Rice picture \cite{BR71} of the IMT. We note that this divergence coincides with the drop of the spectral weight for the $t_{2g}$ and $e_g$ orbitals at the Fermi level as shown in Fig. \ref{Fig_3}.
Our findings clearly indicate the crucial importance of electronic correlations to explain the electronic structure and lattice properties of correlated transition metal oxides.

%%%%%%%%%%%%%%%%%%%%%%%%%%%%%%%%%%%%%%%%
% On the mechanism of magnetic collapse
%%%%%%%%%%%%%%%%%%%%%%%%%%%%%%%%%%%%%%%%

Overall, our calculations reveal that the pressure-induced Mott IMT is accompanied by a simultaneous local moment collapse and by a discontinuous change of the lattice volume. Moreover, under pressure all these materials exhibit a continuous transformation from a localized to itinerant moment behavior of the $3d$ electrons. 
This is in line with the classical case of Mott transition, in which IMT concurs with a complete collapse of magnetism.
We note that magnetic collapse is also found to occur in the paramagnetic B1 phase of NiO, which may have important implications for understanding of the actual driving force behind the magnetic collapse.
In fact, the Ni$^{2+}$ ion has a 3$d^8$ electron configuration with completely occupied $t_{2g}$ and half-filled $e_g$ bands. Therefore changes in the crystal field splitting with pressure cannot result in a magnetic collapse transition in NiO.
On the basis of these observations, we suggest that it is primary a localized to itinerant moment behavior transition which occurs at the IMT that gives rise to magnetic collapse in transition metal oxides. The crystal field however governs the character of the magnetic collapse as well as the magnitude of the high- and low- spin moments at a given pressure.
%
%%%%%%%%%%%%%%%%%%%%%%%%%%%%%%%%%%%%%
% Potential geophysical implications
%%%%%%%%%%%%%%%%%%%%%%%%%%%%%%%%%%%%%
%
Moreover, our results for the magnetic collapse and discontinuous change of the lattice volume found in these compounds at high pressures may have important geophysical implications.
In fact, these phenomena may affect the structure and seismic properties of the Earth interior.
%
%Thus, the magnetic collapse transition obtained in our calculations range from about 40 GPa for CoO to 429 GPa for NiO, i.e., are relevant for the Earth's lower mantle.
%
%It is now clear that extrapolation of low-pressure and temperature results, as well as application of standard band-structure methods, which neglect the effect of strong correlations of localized $3d$ electrons, can lead to inadequate description of the electronic and structural properties. This therefore can dramatically alter our understanding of the phase stability of the Earth's interior, which may require a partial revision of the most accepted geophysical models.
%

In conclusion, we determine the electronic structure and phase stability of paramagnetic B1-structured transition metal oxides at high pressure and temperature.
Our results reveal a pressure-induced Mott insulator-metal transition which is accompanied by a simultaneous collapse of local moments and lattice volume. Upon compression, we observe a continuous transformation from a localized to itinerant moment behavior of the $3d$ electrons. We argue that the magnetic collapse is primary due to a localized to itinerant moment behavior transition which occurs at the IMT.
We point out the importance of further theoretical and experimental investigations of the behavior of transition metal oxides at the high-pressure and temperature range for better understanding of the Earth's interior.

%\begin{acknowledgments}
We thank V. I. Anisimov and D. Vollhardt for valuable discussions. I.L. acknowledges support by the Deutsche Forschungsgemeinschaft through Transregio TRR 80 and the Ministry of Education and Science of the Russian Federation in the framework of Increase Competitiveness Program of NUST 'MISIS' (K3-2016-027). L.P. acknowledges the financial support of the Ministry of Education and Science of the Russian Federation in the framework of Increase Competitiveness Program of NUST 'MISIS' (K3-2015-038) and computational resources provided by the Swedish National Infrastructure for Computing (SNIC) at the National Supercomputer Centre (NSC) and PDC Center for High Performance Computing. A.G. acknowledges the support of the European Research Council (ERC-319286 QMAC) and of the Swiss National Science Foundation (NCCR MARVEL). I.A.A. acknowledges the support from the Swedish Research Council (VR) grant No. 2015-04391, the Swedish Foundation for Strategic Research (SSF) grant SRL 10-0026, the Knut and Alice Wallenberg Foundation through Grant No. 2014-2019 and the Swedish Government Strategic Research Area Grant Swedish e-Science Research Centre (SeRC). The support from Ministry of Education and Science of the Russian Federation (Grant No. 14.Y26.31.0005) is gratefully acknowledged. 

%\end{acknowledgments}


\begin{thebibliography}{99}
\bibitem{different_U_J}
We note however that the calculated transition pressure $P_c$ depends very sensitively on the choice of the interaction parameters $U$ and $J$. In particular, for MnO, $P_c \sim 132.9$ GPa for $U=7$ eV and $J=0.86$ eV and is about $108.8$ GPa for $U=8$ eV and $J=0.75$ eV, respectively. For FeO, it is about $55$ GPa for $U=5$ eV and $J=0.89$ eV, and $\sim 80$ GPa for $U=8$ eV and $J=0.89$ eV; for NiO, $\sim 248$ GPa for $U=8$ eV and $J=1$ eV.
\end{thebibliography}
\end{document}